\documentclass{article}

\PassOptionsToPackage{numbers,sort&compress}{natbib}
\usepackage[preprint]{neurips_2026}

\usepackage[utf8]{inputenc}
\usepackage[T1]{fontenc}
\usepackage{hyperref}
\usepackage{url}
\usepackage{booktabs}
\usepackage{amsfonts}
\usepackage{nicefrac}
\usepackage{microtype}
\usepackage{xcolor}
\usepackage{graphicx}
\usepackage{subcaption}
\usepackage{amsmath}
\usepackage{amssymb}
\usepackage{algorithm}
\usepackage{algorithmic}
\usepackage{wrapfig}
\usepackage{etoolbox}
\usepackage{placeins}

\graphicspath{{figures/}}

\title{X-AuT: Progressive Audio-Encoder Compression for Speech LLMs with Cross-Scale Distillation}

\author{%
  {\bfseries Haojun Zhang, Yi Zou$^\dagger$, Min Chen, Qize Yu, Lianrui Fan} \\[4pt]
  {\bfseries Xini Ding, Hao Li, Shuchang Zhou, Xianming Liu, Shiyu Huang$^\ddagger$} \\[8pt]
  {\normalfont\small XPeng Inc.} \\[5pt]
  {\normalfont\footnotesize\texttt{\{zhanghj15,zouy12,chenm36,yuqz,fanlr1,dingxn2\}@xiaopeng.com}} \\[3pt]
  {\normalfont\footnotesize\texttt{\{lih87,zhousc6,xianming.liu,huangsy16\}@xiaopeng.com}}
}

\begin{document}

\maketitle
\begingroup
\renewcommand{\thefootnote}{\fnsymbol{footnote}}
\footnotetext[2]{Corresponding author.}
\footnotetext[3]{Project leader.}
\endgroup

\begin{abstract}
Reducing audio-encoder depth lowers the inference cost of speech large language models, but removing complete blocks perturbs the embeddings consumed by the decoder and can cause deletion and premature end-of-sequence errors. We introduce X-AuT, a progressive framework that selects layer combinations through short behavioral probes and restores the pruned model through representation alignment, cross-scale distillation, scheduled student-policy supervision, and LoRA finetuning. The language-model backbone remains frozen, while attention LoRA adapters and the tied output embedding adapt during distillation. Training uses the highest-agreement tier from a transcript-consistency pipeline, followed by source reweighting during finetuning. On ten public Chinese--English benchmarks, compressing Qwen3-ASR-0.6B from 18 to 16 audio-encoder layers reduces macro-average error from 5.61\% to 5.27\%. The 14-layer model reaches 5.75\% with 20.7\% fewer audio-tower parameters. Under the matched recipe, the 1.7B teacher yields 5.55\% mean error, compared with 8.45\% for self-distillation, and progressive 18$\rightarrow$14 pruning outperforms direct pruning (5.75\% vs. 6.73\%). These single-run results establish two practical operating points and show that the accuracy effects vary across benchmarks.
Project website: \url{https://xpeng-ai.github.io/x-aut}.
\end{abstract}

\section{Introduction}

Speech large language models typically combine a deep audio encoder, a bridge that maps acoustic features into the language-model embedding space, and an autoregressive text decoder~\cite{qwen3asr,slamasr,whisper}. This architecture is accurate and flexible, but the audio encoder must process every input frame and therefore remains important for first-token latency in streaming, mobile, and in-vehicle systems~\cite{he2019streaming,han2016deep}. Reducing encoder depth is attractive because it removes complete Transformer blocks and produces a regular, deployment-friendly model.

Starting from a strong pretrained model is also substantially cheaper than training a compact speech encoder and realigning it with a decoder from scratch. We therefore study post-training depth reduction of the 18-layer audio Transformer (AuT) in Qwen3-ASR-0.6B~\cite{qwen3asr}. Removing layers changes the audio embeddings inserted into the decoder and can trigger premature end-of-sequence (EOS) predictions and large deletion errors. Our recipe freezes the pretrained language-model weights while allowing decoder attention LoRA adapters and, during distillation, the tied output embedding to adapt. We use ``frozen decoder backbone'' in this sense throughout the paper.

Prior ASR compression work has explored decoder distillation, low-rank encoder compression, weight sparsity, supernet training, and encoder-layer removal~\cite{distilwhisper,liteasr,structuredsparsity,dynamicencoder,encoderrole}. For an already-pretrained speech LLM, two practical questions remain: which combinations of layers can be removed and recovered under a fixed training budget, and how should recovery address both hidden-state mismatch and errors induced by the student's own decoding history?

These questions are coupled. A layer that appears redundant in isolation may become important once another layer is removed, because subsequent blocks receive a shifted representation and the bridge must preserve the decoder interface learned during pretraining. Static importance scores therefore cannot fully predict whether a multi-layer candidate can be recovered within a short budget. Selection and recovery must therefore be evaluated together.

X-AuT addresses these questions through progressive pruning and recovery. Before each pruning hop, short behavioral probes compare candidate layer sets under matched initialization, data, and optimization. The selected student then undergoes representation alignment, distillation with scheduled student-policy contexts, and low-rate LoRA finetuning. A Qwen3-ASR-1.7B teacher supplies cross-scale supervision; grouped layer matching and a learned bottleneck projection accommodate the depth and width differences between teacher and student.

\begin{wrapfigure}{r}{0.49\textwidth}
    \vspace{-8pt}
    \centering
    \includegraphics[width=0.98\linewidth]{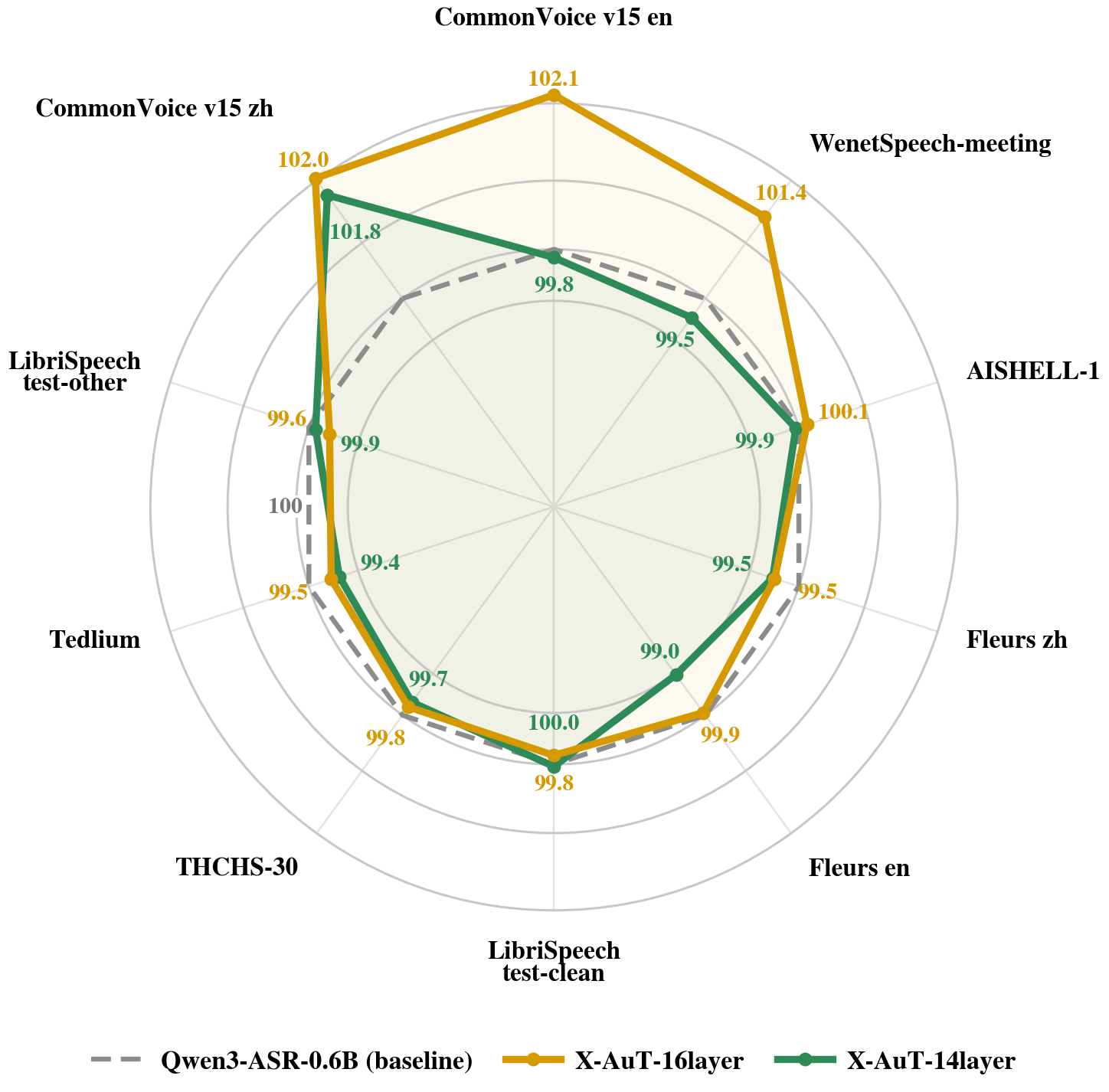}
    \caption{Per-benchmark accuracy preservation of the Stage~2 X-AuT models relative to unpruned Qwen3-ASR-0.6B. The dashed contour marks the baseline (100); higher values indicate better accuracy retention.}
    \label{fig:radar}
    \vspace{-8pt}
\end{wrapfigure}
Figure~\ref{fig:radar} compares the Stage~2 16- and 14-layer models with the unpruned baseline across all ten evaluation sets. Each spoke reports accuracy preservation, $100(1-e_{\mathrm{model}})/(1-e_{\mathrm{base}})$, where $e$ is CER or WER expressed as a fraction. This transformation provides a common visual reference across benchmarks; all quantitative comparisons use the macro-average error and the exact CER/WER values in Tables~\ref{tab:main_full} and~\ref{tab:main_full_14}.

The two contours show different accuracy--efficiency tradeoffs. The 16-layer model remains close to or above the baseline across the suite and reaches 5.27\% macro error. Further compression to 14 layers yields 5.75\% macro error while reducing audio-tower parameters by 20.7\% (186.4M$\rightarrow$147.8M), with most of the additional loss concentrated on a few English benchmarks.

\begin{figure}[!htb]
    \centering
    \includegraphics[width=\textwidth]{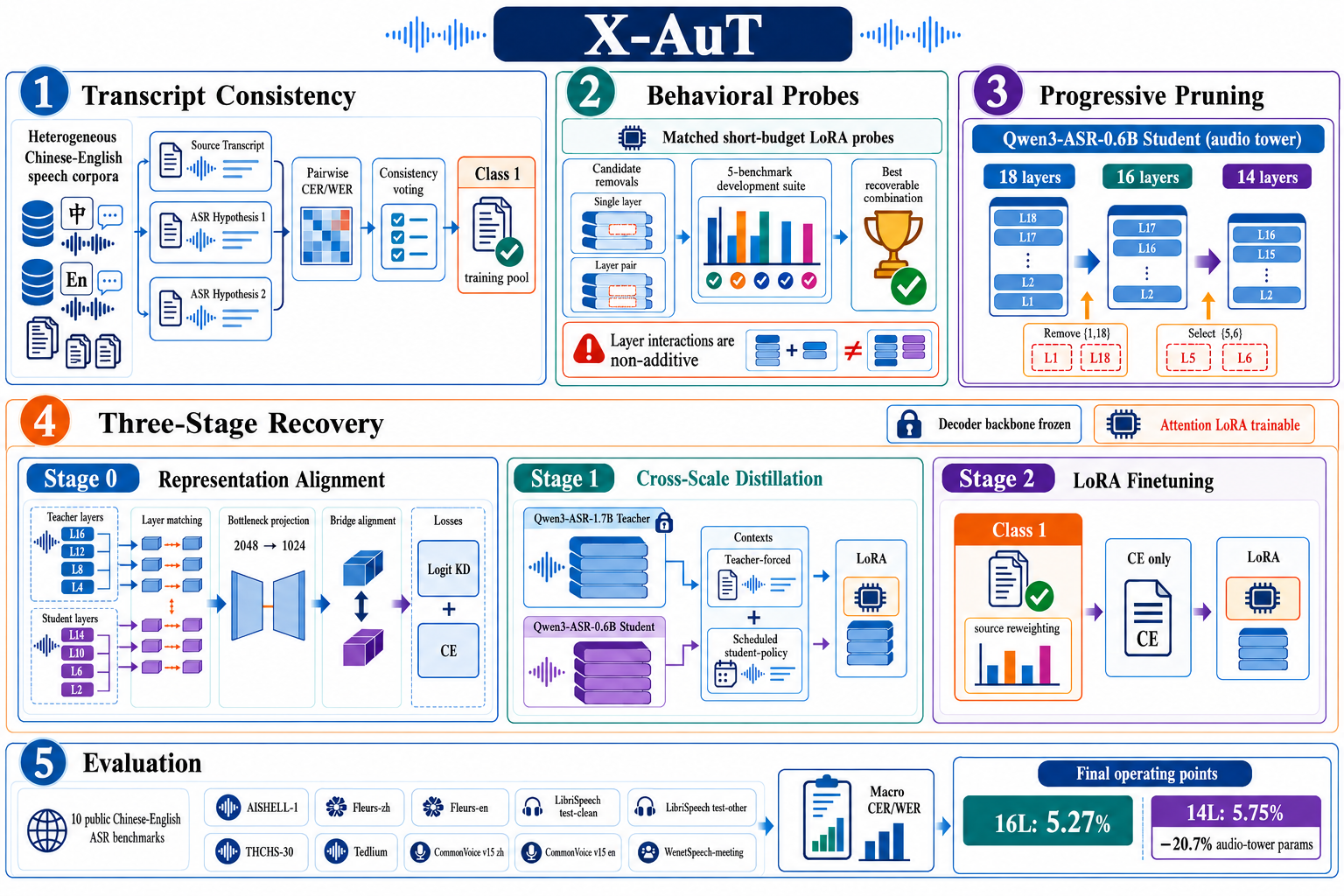}
    \caption{Overview of X-AuT. Transcript-consistency filtering constructs the class~1 training pool, matched short-budget probes select recoverable layer combinations, and two pruning hops reduce the audio tower from 18 to 14 layers. Each pruned student is recovered through representation alignment, cross-scale distillation with teacher-forced and scheduled student-policy contexts, and LoRA finetuning while the decoder backbone remains frozen. Evaluation aggregates CER/WER over ten benchmarks; Secs.~\ref{sec:data} and~\ref{sec:recovery} provide the full configuration.}
    \label{fig:pipeline}
\end{figure}

The benchmark variation also shows why target depth alone is not enough: layer combinations differ in recoverability, and the pruned encoder must be realigned with the decoder under both teacher-forced and student-generated contexts. Figure~\ref{fig:pipeline} connects the main steps, from transcript-consistency filtering and behavioral probes to progressive pruning and three-stage recovery. The reported configuration uses class~1 data in all three stages, with source reweighting during Stage~2.

Our main contributions are:

\begin{itemize}
    \item We introduce a progressive recovery framework that combines behavioral candidate screening, cross-scale hidden-state and logit supervision, scheduled student-policy training, and LoRA finetuning while preserving the pretrained decoder backbone.
    \item A matched teacher-scale comparison reduces mean error from 8.45\% with self-distillation to 5.55\% with the 1.7B teacher, demonstrating the practical value of cross-scale supervision in the tested setting.
    \item Layer-pair probes expose non-additive interactions: the $\{6,8\}$ pair combines the two strongest single removals but underperforms $\{5,6\}$ by 0.85 pp after matched recovery.
    \item The 14-layer model reaches 5.75\% macro error, compared with 5.61\% for the baseline, while removing 20.7\% of audio-tower parameters and reducing measured encoder latency by 21.4\% on the in-vehicle accelerator and 11.4\% on H800.
\end{itemize}
\FloatBarrier

\section{Related Work}

\subsection{Speech Architectures and Encoder Compression}

Modern speech systems increasingly couple an acoustic encoder with a pretrained language model. Qwen3-ASR~\cite{qwen3asr} inserts bridged audio representations into the Qwen3 decoder input, while SLAM-ASR~\cite{slamasr} examines lightweight connectors between speech encoders and LLMs. Qwen2-Audio~\cite{qwen2audio} similarly integrates acoustic representations with a general-purpose language model. Whisper~\cite{whisper} follows an encoder--decoder architecture rather than an LLM-connector design, but remains an important reference for multilingual ASR and subsequent compression work. In each case, the encoder processes the full acoustic sequence, making its depth a direct contributor to computation and latency.

ASR compression has targeted different parts of this architecture. Distil-Whisper~\cite{distilwhisper} primarily reduces the decoder while retaining the Whisper encoder. LiteASR~\cite{liteasr} combines low-rank factorization with distillation for encoder matrices, whereas structured sparsity removes weights or attention heads~\cite{structuredsparsity}. LayerDrop~\cite{layerdrop} trains networks to tolerate variable depth, and Dynamic Encoder Size~\cite{dynamicencoder} learns a supernet from which multiple encoder depths can be extracted. These approaches either introduce compression during training or reduce computation within existing blocks. Direct removal of complete blocks offers a regular architecture, but it also changes the representations consumed by downstream modules.

Kolluri et al.~\cite{encoderrole} study Whisper encoder-layer pruning in an LLM-based SLAM-ASR model and recover the pruned network with LoRA~\cite{lora}. X-AuT instead starts from a pretrained Qwen3-ASR audio tower, removes layers in successive hops, and uses short post-removal probes to compare candidate layer sets under a shared recovery budget. This design treats recoverability as a property of a layer combination rather than an additive score assigned to individual layers.

\subsection{Distillation and Data Selection}

Knowledge distillation can transfer teacher output distributions, intermediate representations, or both. Teacher-forced logit distillation evaluates the teacher and student under gold-prefix contexts, which is efficient but differs from inference once the student conditions on its own predictions. On-policy distillation instead supplies supervision on student-generated histories~\cite{opdsurvey}. Ark-ASR~\cite{arkasr} develops data-efficient on-policy distillation for ASR, and ASKD-Whisper~\cite{askdwhisper} adjusts the strength of self-distillation during training. X-AuT combines the two regimes: teacher-forced supervision remains the default, while scheduled student-policy batches expose the teacher to the student's decoding context after an initial stabilization period. Rollout filtering returns degenerate batches to the teacher-forced objective.

Training data also shape recovery. Curriculum and data-selection methods rank, order, or filter examples according to estimated learning value or label reliability~\cite{curriculumscoring}. Our preprocessing compares the source transcript with hypotheses from two external ASR systems and assigns a transcript-consistency tier from their pairwise agreement. The reported experiments use the highest-agreement tier throughout recovery and alter source weights during Stage~2 to emphasize target-domain data. The resulting pipeline combines confidence-based filtering with phase-specific resampling without relying on a staged mixture of progressively noisier tiers.

\section{Method}

X-AuT filters training examples by transcript agreement, identifies recoverable layer combinations through short behavioral probes, and restores each progressively pruned model with a three-stage training schedule (Figure~\ref{fig:pipeline}).

\subsection{Architecture and Objective}
\label{sec:prelim}

An input waveform $\mathbf{x}$ is encoded by an $N$-layer audio encoder $E_\theta$ and bridge $B_\phi$ into audio embeddings $\mathbf{e}=B_\phi(E_\theta(\mathbf{x}))$. Qwen3-ASR places these embeddings at audio-placeholder positions in the token embedding sequence; the causal language-model decoder $D_\psi$ then predicts transcription tokens through a tied output projection $H_\omega$. This is input-embedding conditioning rather than a separate decoder cross-attention module.

A pruning operation retains an ordered subset $\mathcal{I}\subset\{1,\ldots,N\}$ and forms $E_{\theta,\mathcal{I}}$ from those blocks. We seek a recoverable subset and parameters that minimize aggregate text error rate (TER) under a target depth:
\begin{equation}
  \min_{\mathcal{I},\theta',\phi',\omega'} \operatorname{TER}(E_{\theta',\mathcal{I}},B_{\phi'},D_\psi,H_{\omega'})
  \quad\text{s.t.}\quad |\mathcal{I}|=M<N.
\end{equation}
The pretrained weights of $D_\psi$ remain frozen. LoRA parameters attached to its q/k/v/o attention projections are trainable, and $H_\omega$---which shares weights with the token embedding in Qwen3-ASR---is trainable during Stages~0--1 and frozen during Stage~2.

Removing encoder blocks changes the conditioning embeddings presented to the decoder. The recovery objective therefore first aligns intermediate and bridge representations, then adapts token distributions under teacher-forced and student-generated prefixes.

\subsection{Transcript-Consistency Filtering}
\label{sec:curriculum}

The source pool contains heterogeneous supervision. For each reference-bearing utterance, two strong ASR systems produce offline hypotheses. After language-aware normalization, we compute the three pairwise edit rates among the source transcript and the two hypotheses, using CER for Chinese and WER for English. Their maximum, $e_{\mathrm{max}}$, measures the largest disagreement within the transcript--hypothesis triplet. Exact agreement, Mandarin homophone agreement, and a consistency vote assign one of the nine tiers in Table~\ref{tab:data_quality_classes}. Lower tier numbers indicate stronger transcript agreement rather than ground-truth quality.

\begin{table}[!htb]
    \caption{Nine-tier transcript-consistency hierarchy. Threshold bands use $e_{\mathrm{max}}$, the maximum pairwise edit rate. The tiers measure transcript agreement rather than ground-truth quality.}
    \label{tab:data_quality_classes}
    \centering
    \small
    \begin{tabular}{@{}c p{0.24\linewidth} p{0.57\linewidth}@{}}
        \toprule
        \textbf{Tier} & \textbf{Agreement band} & \textbf{Assignment rule} \\
        \midrule
        1 & Full agreement & Both model hypotheses exactly match the source transcript. \\
        2 & Partial exact & At least one of the three text pairs matches exactly (i.e., at least two candidates agree). \\
        3 & Homophone & For Mandarin, at least one source--hypothesis pair has zero pinyin CER. \\
        4 & \mbox{$0<e_{\mathrm{max}}\leq5\%$} & The consistency vote passes and $e_{\mathrm{max}}$ falls in the indicated interval. \\
        5 & \mbox{$5\%<e_{\mathrm{max}}\leq10\%$} & Same voting rule with a small transcript discrepancy. \\
        6 & \mbox{$10\%<e_{\mathrm{max}}\leq20\%$} & Same voting rule with a moderate discrepancy. \\
        7 & \mbox{$20\%<e_{\mathrm{max}}\leq30\%$} & Same voting rule with a relatively large discrepancy. \\
        8 & \mbox{$30\%<e_{\mathrm{max}}\leq50\%$} & Same voting rule with a large discrepancy. \\
        9 & Failure / \mbox{$e_{\mathrm{max}}>50\%$} & The vote fails, $e_{\mathrm{max}}>50\%$, or an aligned hypothesis is empty. \\
        \bottomrule
    \end{tabular}
\end{table}

The reported configuration uses class~1 for both distillation stages. Stage~2 retains the same consistency threshold but reweights sources toward cockpit and other target-domain data, separating confidence-based filtering from phase-specific source sampling.

\subsection{Behavior-Driven Progressive Pruning}
\label{sec:probe}

We prune in two hops, 18$\rightarrow$16$\rightarrow$14. The first hop removes original layers $\{1,18\}$. For the second hop, every candidate starts from the same recovered 16-layer checkpoint and is trained with the same 0.3-epoch LoRA warm-up. We first evaluate each remaining layer as a single removal, then evaluate a fixed set of adjacent and non-adjacent layer pairs. Candidate selection uses the lowest aggregate TER on a fixed five-benchmark development suite (Sec.~\ref{sec:expsetup}). This procedure is more expensive than a static score but directly measures post-removal behavior under the available recovery budget.

The pair probes are necessary because recovery after removing several layers cannot be predicted reliably from the corresponding single-layer scores. The matched comparison selects $\{5,6\}$ for the 16$\rightarrow$14 hop; Section~\ref{sec:interaction} reports the candidate-level results.

\subsection{Three-Stage Recovery}
\label{sec:recovery}

Each hop uses the same three-stage recipe. The student is the pruned Qwen3-ASR-0.6B model; the teacher is Qwen3-ASR-1.7B with a 24-layer audio encoder. Teacher parameters are frozen and discarded at inference.

\subsubsection{Stage 0: Representation Alignment}

Stage~0 occupies the first 5\% of the distillation epoch and combines intermediate-layer, bridge, logit, and transcript losses:
\begin{equation}
\mathcal{L}_{\mathrm{S0}}=
\lambda_{\mathrm{layer}}\mathcal{L}_{\mathrm{layer}}+
\lambda_{\mathrm{bridge}}\mathcal{L}_{\mathrm{bridge}}+
\lambda_{\mathrm{logit}}\mathcal{L}_{\mathrm{logit}}+
\lambda_{\mathrm{ce}}\mathcal{L}_{\mathrm{ce}}.
\end{equation}
Both representation losses sum mean-squared error and cosine distance. Teacher layers are divided uniformly into $M$ ordered groups, and student layer $m$ aligns to the last teacher layer in group $m$. Because the teacher and student hidden widths are 2048 and 1024, respectively, a learned two-layer MLP with a 256-dimensional bottleneck projects teacher hidden and bridge features into the student space. Logit KD uses temperature-scaled KL divergence under gold prefixes.

The pruned audio encoder and bridge are fully trainable. Decoder base weights remain frozen, while rank-32 LoRA adapters on q/k/v/o projections and the tied output embedding are trained with a separate decoder learning rate.

\subsubsection{Stage 1: Distillation with Scheduled Student-Policy Contexts}

Stage~1 disables intermediate-layer loss and uses bridge alignment, teacher-forced logit KD, and gold-transcript CE:
\begin{equation}
\mathcal{L}_{\mathrm{off}}=
\lambda_{\mathrm{bridge}}\mathcal{L}_{\mathrm{bridge}}+
\lambda_{\mathrm{logit}}\mathcal{L}_{\mathrm{logit}}+
\lambda_{\mathrm{ce}}\mathcal{L}_{\mathrm{ce}}.
\end{equation}
After 20\% of Stage~1 has elapsed, every fifth optimizer step is scheduled for student-policy supervision. The student greedily generates a prefix; student and teacher are then evaluated on the same generated context. Their distributions are compared over the union of each model's top-$k$ support ($k=512$). Gold CE remains an anchor on these scheduled steps. The implementation uses no confidence reweighting (\texttt{weight\_mode=none}).

Rollout safeguards prevent degenerate prefixes from entering the KD loss. Generation enforces \texttt{min\_new\_tokens}$=3$, uses a duration-aware maximum capped at 256 tokens, and rejects budget-exhausted, over-long, or repetitive rollouts. If at least half of a batch is rejected, that microbatch falls back to the teacher-forced objective. Thus, approximately 20\% is a scheduling target; the realized on-policy fraction can be lower after filtering.

\subsubsection{Stage 2: LoRA Finetuning}

Stage~2 initializes from the best Stage~1 checkpoint and optimizes gold-transcript CE for one epoch. The audio encoder, bridge, and decoder LoRA adapters remain trainable at $5\times10^{-6}$; the tied lm\_head/embedding is frozen. The class-1 data index is reweighted toward target-domain sources. This stage contains no teacher loss:
\begin{equation}
\mathcal{L}_{\mathrm{S2}}=\mathcal{L}_{\mathrm{ce}}(\mathbf{y}^{\mathrm{gold}},\hat{\mathbf{y}}).
\end{equation}

\section{Experimental Setup}
\label{sec:expsetup}

\subsection{Models and Parameter Accounting}

The student starts from Qwen3-ASR-0.6B~\cite{qwen3asr}, whose audio tower contains 18 Transformer blocks and a convolutional/bridge frontend. Counting tensors in the released checkpoint gives 186.376M audio-tower parameters: 12.758M outside the Transformer stack and 9.645M per block. The 16- and 14-layer students therefore contain 167.085M and 147.794M audio-tower parameters, corresponding to 10.35\% and 20.70\% reductions. We report these exact counts rather than inferring compression from rounded labels such as 180M/140M. The cross-scale teacher is Qwen3-ASR-1.7B with a 24-layer audio tower and 2048-dimensional hidden states; the student audio hidden width is 1024.

\subsection{Training Data and Quality Labels}
\label{sec:data}

The source pool combines public and proprietary multilingual ASR corpora, including AISHELL-1/4/5~\cite{aishell1,aishell4,aishell5}, CommonVoice~\cite{commonvoice}, Emilia~\cite{emilia}, GigaSpeech~\cite{gigaspeech}, KeSpeech~\cite{kespeech}, LibriSpeech~\cite{librispeech}, WenetSpeech~\cite{wenetspeech}, and cockpit-domain speech. The pool exceeds 280k hours before quality filtering. Audio is capped at 40 seconds.

\begin{figure}[!htb]
    \centering
    \includegraphics[width=\linewidth]{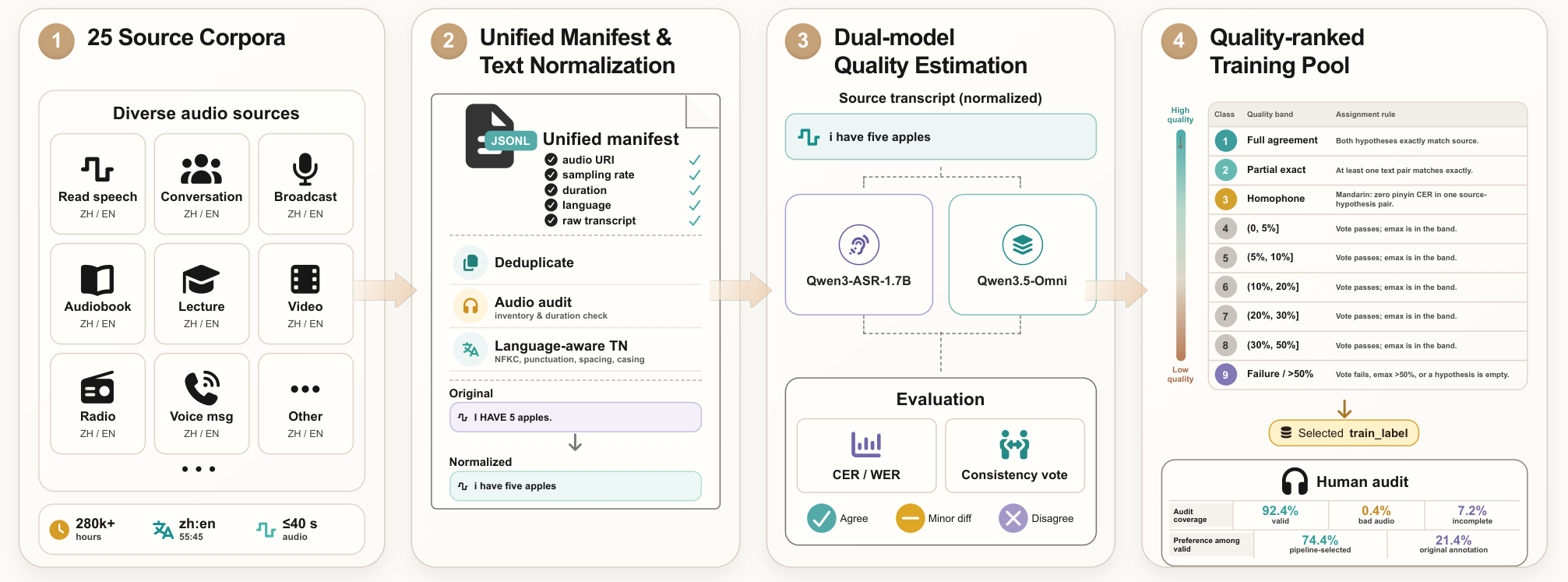}
    \caption{Four-step data pipeline: corpus unification and normalization, dual-system transcription, pairwise CER/WER and consistency voting, and quality-ranked label selection. The rightmost panel reports the WenetSpeech audit summary; audit percentages describe label-selection behavior and are not used as training weights.}
    \label{fig:datapipe}
\end{figure}

\paragraph{Manifest construction.} Each corpus is converted to a unified JSONL manifest containing an utterance identifier, audio reference, source transcript, language and split tags, duration, sampling rate, and channel count. Text normalization includes Unicode NFKC normalization, width conversion, Traditional-to-Simplified conversion for Mandarin, removal of invisible characters and numeric separators, dash canonicalization, and whitespace normalization. Original and normalized text are retained for traceability. Figure~\ref{fig:datapipe} summarizes the complete path from corpus ingestion through transcript agreement and quality ranking.

Reference-bearing utterances are decoded offline by Qwen3-ASR-1.7B and Qwen3.5-Omni~\cite{qwen35omni}. Records without an inference result, a valid audio reference, or nonempty supervision are excluded. The remaining records receive the consistency labels described in Sec.~\ref{sec:curriculum}. The reported distillation runs use the class-1 manifest index; the first-hop log contains approximately 299k weighted target records per epoch and the second-hop log approximately 292k. Stage~2 keeps class~1 but changes corpus weights, increasing AISHELL-4/5 and cockpit-query contributions while dropping several weakly matched web-speech sources. These counts describe the realized loader indices rather than the size of the 280k-hour source pool.

\subsection{Selection and Evaluation Suites}

\paragraph{Development selection suite.} Behavior probes and checkpoint selection use five fixed development or validation subsets: AISHELL-1 (Mandarin CER), CommonVoice-en (English WER), Fleurs-en (English WER), WenetSpeech-meeting (Mandarin CER), and a proprietary cockpit-query subset (Mandarin CER). Frequent evaluation is capped at 25 utterances per benchmark, and the macro average of the five normalized error rates determines checkpoint selection. The final results are evaluated separately on the full public benchmark suite described below.

\paragraph{Full public evaluation suite.} Final checkpoint results are reported on ten public benchmarks: AISHELL-1 (CER)~\cite{aishell1}; Fleurs zh/en (CER/WER)~\cite{fleurs}; LibriSpeech test-clean/test-other (WER)~\cite{librispeech}; THCHS-30 (CER)~\cite{thchs30}; Tedlium (WER)~\cite{tedlium}; CommonVoice v15 zh/en (CER/WER)~\cite{commonvoice}; and WenetSpeech-meeting (CER)~\cite{wenetspeech}. The macro mean weights benchmarks equally, not by utterance count. The proprietary selection subset is excluded from all main-result tables.

\subsection{Optimization and Reporting Protocol}

The reported 0.6B runs use 32 accelerators, per-device batch size 8, gradient accumulation 2, and global batch size 512. We use AdamW with $2\times10^{-5}$ for the audio tower and $1\times10^{-4}$ for decoder-side LoRA plus the tied output embedding during distillation. Stage~2 uses $5\times10^{-6}$ for all trainable parameters and freezes the tied output embedding. Weight decay is 0.01, gradient clipping is 1.0, and training uses bf16. Stage~0 and Stage~1 occupy 0.05 and 0.95 epoch; Stage~2 runs for one epoch. All configurations use seed 42.

Checkpoints are selected by the lowest observed selection-suite macro TER. Tables report the corresponding single-run checkpoint on the full suite. We did not run repeated seeds or bootstrap utterance-level confidence intervals, so boldface denotes the best observed number in a row and not statistical significance. Full hyperparameters are listed in Appendix~\ref{sec:hyperparameters}.

Efficiency is measured separately on an in-vehicle PPU and an NVIDIA H800 GPU using more than 50 utterances of varying duration. We report descriptive averages from the available benchmark output; run-to-run variability was not retained.

\section{Results}

\subsection{Main Results}
\label{sec:benchmarks}

\begin{table}[!htb]
  \caption{Full-suite error rates for the 16-layer model. S1 is the Stage~1 best checkpoint and S2 is the finetuned checkpoint. Mean is the unweighted macro average. Relative mean-error change is $(\bar e/\bar e_{\rm base}-1)\times100\%$; negative is better. Bold marks the best observed value, not statistical significance.}
  \label{tab:main_full}
  \centering
  \small
  \renewcommand{\arraystretch}{1.14}
  \begin{tabular}{@{}lccc@{}}
    \toprule
    Benchmark & Base (18L) & Prune-16 S1 & Prune-16 S2 \\
    AuT parameters & 186.4M & 167.1M & 167.1M \\
    \midrule
    AISHELL-1 (CER) & 3.33\% & 3.30\% & \textbf{3.21\%} \\
    Fleurs-zh (CER) & \textbf{2.80\%} & 3.35\% & 3.28\% \\
    Fleurs-en (WER) & \textbf{4.17\%} & 4.28\% & 4.23\% \\
    LibriSpeech test-clean (WER) & \textbf{2.48\%} & 2.73\% & 2.65\% \\
    THCHS-30 (CER) & \textbf{3.87\%} & 4.10\% & 4.06\% \\
    Tedlium (WER) & \textbf{3.35\%} & 3.92\% & 3.79\% \\
    LibriSpeech test-other (WER) & \textbf{5.39\%} & 5.90\% & 5.79\% \\
    CommonVoice v15 zh (CER) & 9.95\% & 8.56\% & \textbf{8.12\%} \\
    CommonVoice v15 en (WER) & 12.35\% & 10.74\% & \textbf{10.50\%} \\
    WenetSpeech-meeting (CER) & 8.36\% & 8.62\% & \textbf{7.06\%} \\
    \midrule
    \textit{Macro mean (\%)} & 5.61 & 5.55 & \textbf{5.27} \\
    \textit{Relative mean-error change (\%)} & -- & $-1.1$ & \textbf{$-6.1$} \\
    \bottomrule
  \end{tabular}
\end{table}

The 16-layer model lowers macro-average error from 5.61\% to 5.27\%, an absolute change of $-0.34$ pp and a 6.1\% relative error reduction (Table~\ref{tab:main_full}). It improves four benchmarks: AISHELL-1, CommonVoice zh/en, and WenetSpeech-meeting. The largest gains occur on CommonVoice zh ($-1.83$ pp), CommonVoice en ($-1.85$ pp), and WenetSpeech-meeting ($-1.30$ pp), while the largest degradation is 0.44 pp on Tedlium. Stage~2 improves all ten entries relative to the Stage~1 checkpoint.

\begin{table}[!htb]
  \caption{Full-suite error rates for the 14-layer model. The 147.8M audio tower has 20.7\% fewer parameters than the 186.4M baseline. Formatting and reporting conventions follow Table~\ref{tab:main_full}.}
  \label{tab:main_full_14}
  \centering
  \small
  \renewcommand{\arraystretch}{1.14}
  \begin{tabular}{@{}lccc@{}}
    \toprule
    Benchmark & Base (18L) & Prune-14 S1 & Prune-14 S2 \\
    AuT parameters & 186.4M & 147.8M & 147.8M \\
    \midrule
    AISHELL-1 (CER) & \textbf{3.33\%} & 3.52\% & 3.39\% \\
    Fleurs-zh (CER) & \textbf{2.80\%} & 3.49\% & 3.32\% \\
    Fleurs-en (WER) & \textbf{4.17\%} & 5.24\% & 5.10\% \\
    LibriSpeech test-clean (WER) & 2.48\% & 3.09\% & \textbf{2.45\%} \\
    THCHS-30 (CER) & \textbf{3.87\%} & 4.23\% & 4.17\% \\
    Tedlium (WER) & \textbf{3.35\%} & 4.07\% & 3.95\% \\
    LibriSpeech test-other (WER) & \textbf{5.39\%} & 7.00\% & 5.52\% \\
    CommonVoice v15 zh (CER) & 9.95\% & 9.54\% & \textbf{8.36\%} \\
    CommonVoice v15 en (WER) & \textbf{12.35\%} & 13.94\% & 12.49\% \\
    WenetSpeech-meeting (CER) & \textbf{8.36\%} & 10.71\% & 8.78\% \\
    \midrule
    \textit{Macro mean (\%)} & \textbf{5.61} & 6.48 & 5.75 \\
    \textit{Relative mean-error change (\%)} & -- & $+15.5$ & $+2.5$ \\
    \bottomrule
  \end{tabular}
\end{table}

The 14-layer model contains 147.794M audio-tower parameters, 20.70\% fewer than the 186.376M baseline. Its macro error is 5.75\%, a 0.14-pp increase over the baseline (Table~\ref{tab:main_full_14}). CommonVoice zh improves by 1.59 pp and LibriSpeech test-clean by 0.03 pp, whereas Fleurs-en has the largest degradation at 0.93 pp. Relative to the 16-layer model, seven benchmark changes remain within 0.3 pp; CommonVoice en ($+1.99$ pp) and WenetSpeech-meeting ($+1.72$ pp) account for most of the macro-average gap. Figure~\ref{fig:radar} summarizes this benchmark-level variation, and the tables report the corresponding CER/WER values.

The two operating points expose a clear tradeoff. The 16-layer model improves the observed macro average, while the 14-layer model provides a larger structural reduction at a small average cost. Section~\ref{sec:discussion} discusses the uncertainty associated with these single-run comparisons.

\subsection{Training Trajectories}
\label{sec:dynamics}

\begin{figure}[!htb]
    \centering
    \includegraphics[width=0.94\linewidth]{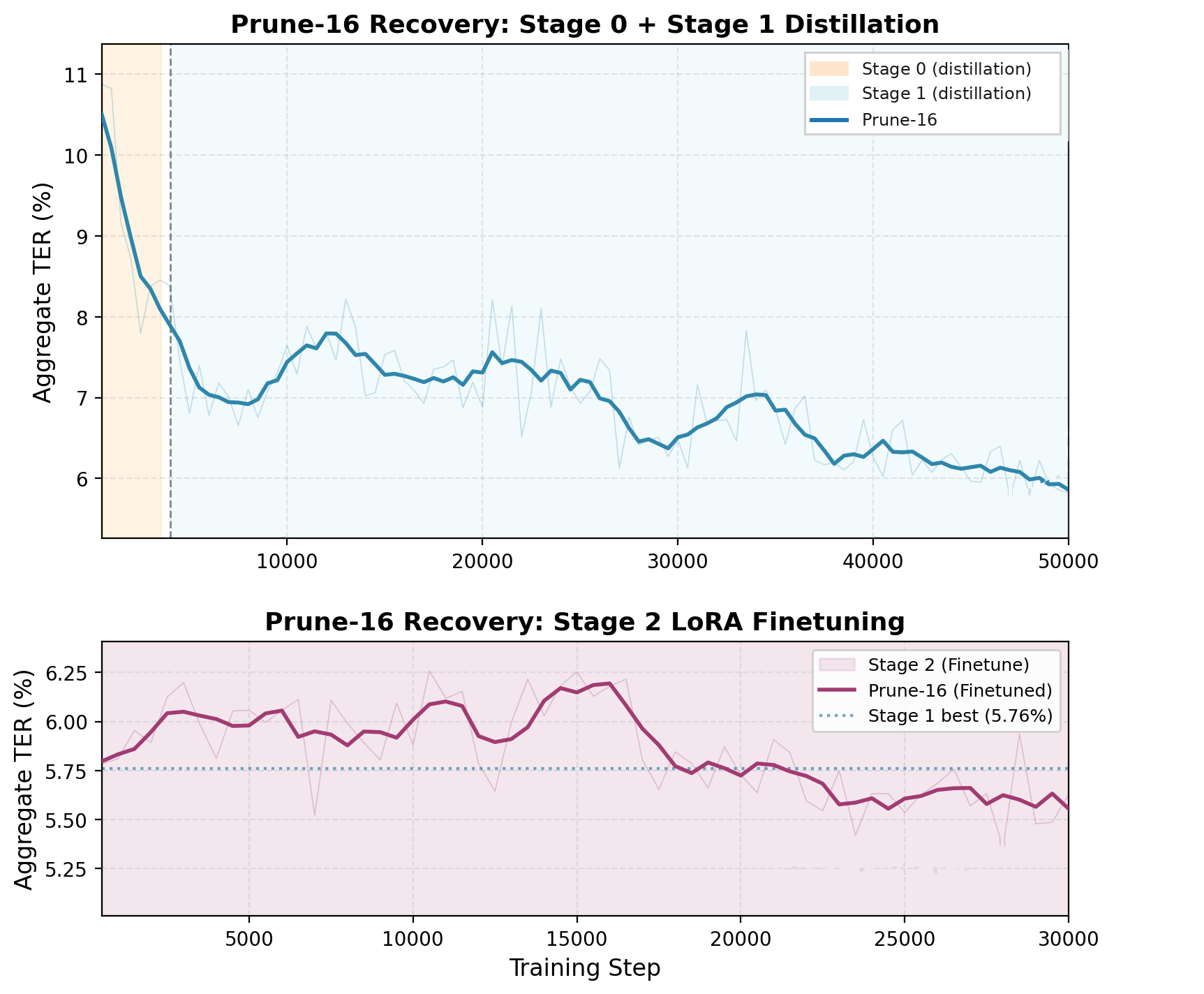}
    \caption{Selection-suite TER during 16-layer recovery. (a) Stage~0/1 distillation, with the transition near step 4000. (b) Stage~2 finetuning from the best Stage~1 checkpoint; the dashed line marks 5.76\%. The displayed trends summarize checkpoint evaluations and are not uncertainty estimates.}
    \label{fig:ter_curve}
\end{figure}

Stage~0 TER falls from 10.88\% at step 500 to 7.80\% at step 2500, followed by 8.38\% at the first Stage~1 evaluation near step 4000 (Figure~\ref{fig:ter_curve}). Stage~1 reaches its minimum of 5.76\% at step 47,000. Starting from that checkpoint, Stage~2 reduces TER by another 0.40 pp and reaches 5.36\% at step 28,000.

The trajectory shows rapid early recovery followed by slower optimization in Stage~1 and a further gain from Stage~2. Because the stage transition changes both the loss and optimizer state, the curve describes the recovery process rather than the contribution of an individual loss term.

\subsection{Teacher Scale}
\label{sec:crossscale}

A same-scale control replaces the 1.7B teacher with the student's unpruned 0.6B model while retaining the Stage~0/1 schedule, data, LoRA configuration, and optimization hyperparameters. The cross-scale case additionally requires the learned 2048$\rightarrow$1024 teacher projections. At the Stage~1 best checkpoints, mean full-suite error is 5.55\% for the cross-scale teacher and 8.45\% for the self-teacher; the cross-scale model is better on all ten benchmarks (Appendix~\ref{sec:teacher_ablation}). Relative to the unpruned baseline, the cross-scale checkpoint improves macro error by 1.1\%, while the self-teacher checkpoint increases it by 50.6\%.

The large gap demonstrates the practical value of the stronger teacher under the implemented recipe. On CommonVoice zh/en, the cross-scale checkpoint also improves over the original 0.6B baseline, suggesting that the larger teacher transfers useful acoustic behavior rather than merely restoring the pruned student to its starting point. The scope of this interpretation is discussed in Section~\ref{sec:discussion}.

\subsection{Behavior-Driven Layer Selection}
\label{sec:interaction}

\begin{figure}[!htb]
    \centering
    \includegraphics[width=0.98\textwidth]{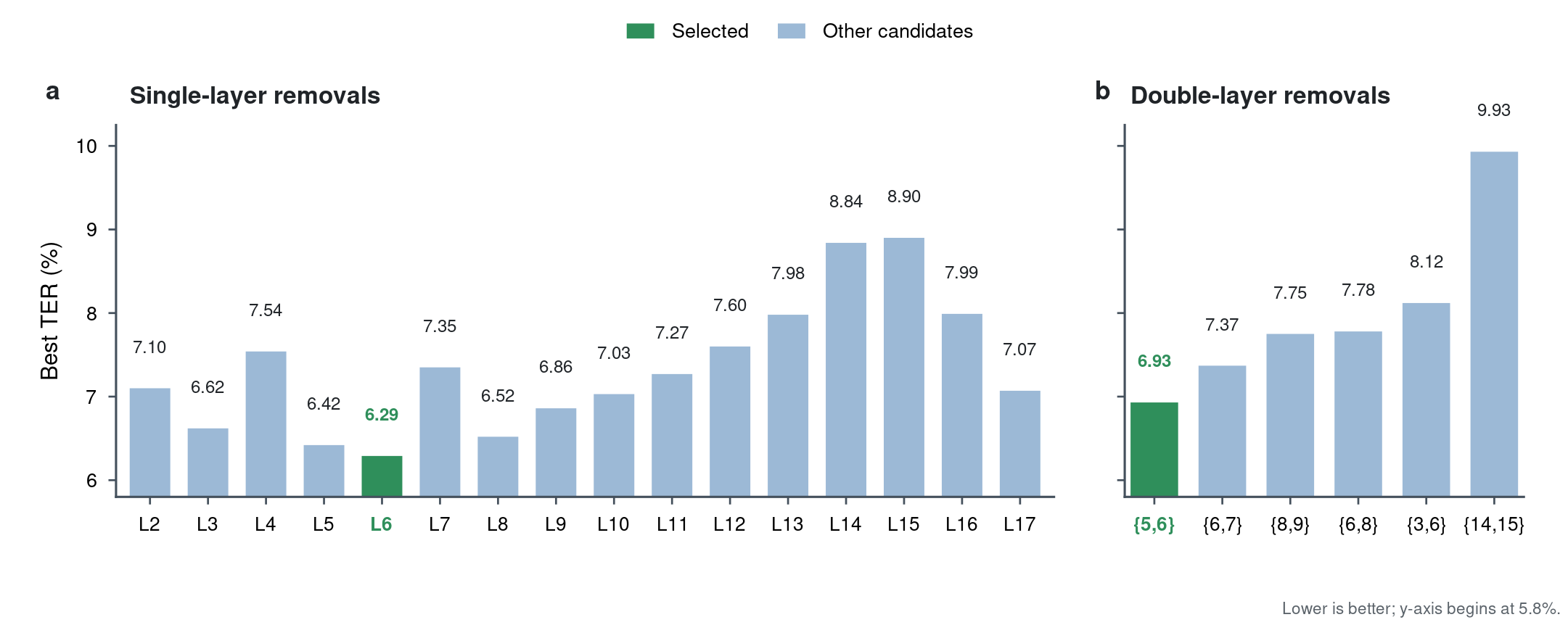}
    \caption{Layer-screening results for single- and double-layer removals after the same 0.3-epoch warm-up. Lower TER is better; green marks the candidate selected for progressive pruning.}
    \label{fig:probe_merged}
\end{figure}

The single-layer sweep spans 6.29--8.90\% TER. L6 is the strongest single removal at 6.29\%, followed by L5 at 6.42\% and L8 at 6.52\%. The selected $\{5,6\}$ pair reaches 6.93\%, whereas $\{6,8\}$ reaches 7.78\%. Defining the interaction penalty as pair TER minus the mean TER of its constituent removals gives 0.58 pp for $\{5,6\}$ and 1.38 pp for $\{6,8\}$.

The adjacent candidates $\{5,6\}$, $\{6,7\}$, and $\{8,9\}$ rank ahead of the tested non-adjacent candidates $\{6,8\}$ and $\{3,6\}$, while the adjacent $\{14,15\}$ pair performs worst overall. Thus, adjacency alone is not a reliable selection rule, and pair recoverability cannot be inferred from single-layer scores in isolation.

\section{Discussion}
\label{sec:discussion}

\subsection{Interpretation}

Taken together, the results suggest that successful depth reduction depends on both representation recovery and retained encoder capacity. Removing layers perturbs the audio embeddings presented to the decoder, as reflected in the early TER trajectory and the EOS diagnostics. Stage~0 directly reduces this mismatch at the intermediate and bridge levels, and Stage~1 extends recovery to token distributions under gold and student-generated prefixes. This procedure is sufficient for the 16-layer model to surpass the baseline macro error, but the remaining gap at 14 layers indicates that alignment cannot fully replace the capacity lost through deeper pruning.

The teacher-scale comparison further shows that the source of the recovery signal matters. Under the matched 16-layer recipe, the 1.7B teacher yields 5.55\% mean error, compared with 8.45\% for self-distillation from the unpruned 0.6B model. The improvement spans all ten benchmarks and is particularly pronounced on CommonVoice zh/en. We interpret this result as useful transfer from the larger teacher within the present training recipe, while recognizing that the comparison does not separate pruning recovery from gains that the same recipe might provide to an unpruned student.

Layer choice and pruning schedule also affect recoverability. Although L8 and L6 are the two strongest single-layer removals, pruning them together performs worse than the selected $\{5,6\}$ pair after matched recovery. The pair probes therefore provide information that cannot be inferred by ranking individual layers alone. Similarly, progressive 18$\rightarrow$16$\rightarrow$14 pruning reaches 5.75\% mean error, whereas direct 18$\rightarrow$14 pruning reaches 6.73\% under the same nominal budget. These comparisons favor explicit pair evaluation and progressive pruning for the model and candidates studied here.

\subsection{Limitations}

The primary results are single runs with seed 42, without repeated-seed variation, paired utterance-level confidence intervals, or significance tests. Checkpoint selection also relies on fixed development subsets capped at 25 utterances per benchmark. This design makes frequent evaluation tractable, but it introduces selection noise and leaves small differences, including the 0.14-pp gap between the 14-layer model and the baseline, as descriptive observations.

The study covers one model family and a limited set of pruning candidates. Layer interactions, projection-based alignment, and EOS behavior may differ in Whisper~\cite{whisper}, Qwen2-Audio~\cite{qwen2audio}, or other speech--language architectures. The data pipeline likewise supports nine consistency classes, whereas the reported runs use class~1 throughout and change only source weights in Stage~2. Experiments across model families, broader layer combinations, and controlled data mixtures would clarify which parts of the recipe generalize beyond the present setting.

Only the audio tower is compressed, so autoregressive decoding limits the reduction in end-to-end latency. The efficiency measurements are averages from one retained benchmark output and do not include run-to-run uncertainty. A complete deployment study should repeat the hardware measurements under a fixed software stack and report latency distributions alongside average values.

\section{Conclusion}

X-AuT reduces the Qwen3-ASR-0.6B audio tower from 18 to 14 layers through progressive pruning and cross-scale recovery. The 16-layer model improves macro-average error from 5.61\% to 5.27\%, while the 14-layer model reaches 5.75\% with 20.7\% fewer audio-tower parameters. The teacher-scale, pair-selection, and direct-pruning controls show that recovery depends on teacher strength, joint layer selection, and pruning schedule. Because the results come from single runs on one model family, broader models and repeated trials are needed to determine how well this tradeoff generalizes.

\clearpage
\bibliographystyle{unsrtnat}
\bibliography{refs}

@misc{qwen3asr,
  title        = {Qwen3-ASR Technical Report},
  author       = {Xian Shi and Xiong Wang and Zhifang Guo and Yongqi Wang and Pei Zhang and Xinyu Zhang and Zishan Guo and Hongkun Hao and Yu Xi and Baosong Yang and Jin Xu and Jingren Zhou and Junyang Lin},
  year         = {2026},
  eprint       = {2601.21337},
  archiveprefix = {arXiv},
  primaryclass = {cs.CL}
}

@misc{qwen35omni,
  title        = {Qwen3.5-Omni Technical Report},
  author       = {{Qwen Team}},
  year         = {2026},
  eprint       = {2604.15804},
  archiveprefix = {arXiv},
  primaryclass = {cs.CL}
}

@misc{qwen2audio,
  title        = {Qwen2-Audio Technical Report},
  author       = {{Qwen Team}},
  year         = {2024},
  eprint       = {2407.10759},
  archiveprefix = {arXiv}
}

@article{whisper,
  title        = {Robust Speech Recognition via Large-Scale Weak Supervision},
  author       = {Radford, Alec and Kim, Jong Wook and Xu, Tao and Brockman, Greg and McLeavey, Christine and Sutskever, Ilya},
  journal      = {arXiv preprint arXiv:2212.04356},
  year         = {2022}
}

@misc{slamasr,
  title        = {An Embarrassingly Simple Approach for LLM with Strong ASR Capacity},
  author       = {Ziyang Ma and Guanrou Yang and Yifan Yang and Zhifu Gao and Jiaming Wang and Zhihao Du and Fan Yu and Qian Chen and Siqi Zheng and Shiliang Zhang and Xie Chen},
  year         = {2024},
  eprint       = {2402.08846},
  archiveprefix = {arXiv},
  primaryclass = {cs.CL}
}

@inproceedings{distilwhisper,
  title        = {Distil-Whisper: Robust Knowledge Distillation via Large-Scale Pseudo Labelling},
  author       = {Gandhi, Sanchit and von Platen, Patrick and Rush, Alexander M.},
  booktitle    = {arXiv preprint arXiv:2311.00430},
  year         = {2023}
}

@inproceedings{liteasr,
  title        = {LiteASR: Efficient Automatic Speech Recognition with Low-Rank Approximation},
  author       = {Kamahori, Keisuke and Kasai, Jungo and Kojima, Noriyuki and Kasikci, Baris},
  booktitle    = {Proceedings of the 2025 Conference on Empirical Methods in Natural Language Processing (EMNLP)},
  year         = {2025}
}

@misc{structuredsparsity,
  title        = {Structured Sparsity and Weight-adaptive Pruning for Memory and Compute efficient Whisper models},
  author       = {Prasenjit K Mudi and Anshi Sachan and Dahlia Devapriya and Sheetal Kalyani},
  year         = {2025},
  eprint       = {2510.12666},
  archiveprefix = {arXiv},
  primaryclass = {cs.LG}
}

@inproceedings{encoderrole,
  title        = {On the Role of Encoder Depth: Pruning Whisper and LoRA Fine-Tuning in SLAM-ASR},
  author       = {Kolluri, Ganesh Pavan Kartikeya Bharadwaj and Kampouridis, Michael and Shekhar, Ravi},
  booktitle    = {Proceedings of the SPEAKABLE Workshop, LREC},
  year         = {2026}
}

@inproceedings{dynamicencoder,
  title        = {Dynamic Encoder Size Based on Data-Driven Layer-wise Pruning for Speech Recognition},
  author       = {Xu, Jingjing and Beck, Eugen and Yang, Zijian and Schl{\"u}ter, Ralf},
  booktitle    = {Proc. Interspeech},
  pages        = {4563--4567},
  year         = {2024}
}

@misc{arkasr,
  title        = {Data-Efficient On-Policy Distillation for Automatic Speech Recognition},
  author       = {Yu Lin and Yiming Wang and Runyuan Cai and Xiaodong Zeng},
  year         = {2026},
  eprint       = {2605.28139},
  archiveprefix = {arXiv},
  primaryclass = {cs.AI}
}

@misc{askdwhisper,
  title        = {ASKD-Whisper: Adaptive Self-knowledge Distillation for Efficient and Low-Latency Automatic Speech Recognition},
  author       = {Junseok Lee and Nahun Kim and Sangyong Lee and Chang-Jae Chun},
  year         = {2026},
  eprint       = {2601.19919},
  archiveprefix = {arXiv},
  primaryclass = {cs.CL}
}

@misc{opdsurvey,
  title        = {A Survey of On-Policy Distillation for Large Language Models},
  author       = {Mingyang Song and Mao Zheng},
  year         = {2026},
  eprint       = {2604.00626},
  archiveprefix = {arXiv},
  primaryclass = {cs.LG}
}

@misc{curriculumscoring,
  title        = {Does the Definition of Difficulty Matter? Scoring Functions and their Role for Curriculum Learning},
  author       = {Simon Rampp and Manuel Milling and Andreas Triantafyllopoulos and Bj{\"o}rn W. Schuller},
  year         = {2024},
  eprint       = {2411.00973},
  archiveprefix = {arXiv},
  primaryclass = {cs.LG}
}

@inproceedings{he2019streaming,
  title        = {Streaming End-to-End Speech Recognition for Mobile Devices},
  author       = {He, Yanzhang and Sainath, Tara N. and Prabhavalkar, Rohit and McGraw, Ian and Alvarez, Raziel and Zhao, Ding and Rybach, David and Kannan, Anjuli and Wu, Yonghui and Pang, Ruoming and Liang, Qiao and Bhatia, Deepti and Shangguan, Yuan and Li, Bo and Pandey, Golan and Sim, Khe Chai and Bagby, Thomas and Chang, Shuo-Yin and Rao, Kanishka and Gruenstein, Alexander},
  booktitle    = {Proc. IEEE ICASSP},
  pages        = {6381--6385},
  year         = {2019}
}

@inproceedings{han2016deep,
  title        = {Deep Compression: Compressing Deep Neural Networks with Pruning, Trained Quantization and Huffman Coding},
  author       = {Han, Song and Mao, Huizi and Dally, William J.},
  booktitle    = {International Conference on Learning Representations (ICLR)},
  year         = {2016}
}

@inproceedings{lora,
  title        = {LoRA: Low-Rank Adaptation of Large Language Models},
  author       = {Hu, Edward J. and Shen, Yelong and Wallis, Phillip and Allen-Zhu, Zeyuan and Li, Yuanzhi and Wang, Shean and Wang, Lu and Chen, Weizhu},
  booktitle    = {International Conference on Learning Representations (ICLR)},
  year         = {2022}
}

@inproceedings{layerdrop,
  title        = {Reducing Transformer Depth on Demand with Structured Dropout},
  author       = {Fan, Angela and Grave, Edouard and Joulin, Armand},
  booktitle    = {International Conference on Learning Representations (ICLR)},
  year         = {2020}
}

@inproceedings{aishell1,
  title        = {Aishell-1: An Open-Source Mandarin Speech Corpus and a Speech Recognition Baseline},
  author       = {Bu, Huihui and Du, Juntao and Na, Xingyu and Ji, Yanyan and Zheng, Fang},
  booktitle    = {Proc. O-COCOSDA},
  year         = {2017}
}

@misc{aishell4,
  title        = {AISHELL-4: An Open Source Dataset for Speech Recognition in Multi-Party Conference Scenario},
  author       = {Fu, Yihui and Xu, Luyao and Zhai, Yukai and Wang, Yuxiang and Liang, Kun and Li, Yongqiang and Zhang, Shiliang and Yan, Yonghong and Chen, Xie},
  year         = {2021},
  eprint       = {2104.03035},
  archiveprefix = {arXiv},
  primaryclass = {eess.AS}
}

@inproceedings{aishell5,
  title        = {AISHELL-5: Multi-Domain Mandarin Speech Recognition Corpus with Labeled Data of 520 Hours},
  author       = {Shi, Yuhao and Xu, Luyao and Zhang, Shiliang and Yan, Yonghong},
  booktitle    = {Proc. IEEE ICASSP},
  year         = {2023}
}

@inproceedings{commonvoice,
  title        = {Common Voice: A Massively-Multilingual Speech Corpus},
  author       = {Ardila, Rosana and Branson, Megan and Lee, Kelly and Kohler, Michael and Schumann, Rebekah and Sterckx, Laure and Bayron, Juan Diez and Karunanayake, Prasanga and Sanabria, Ram{\'o}n and Baas, Andre and others},
  booktitle    = {Proc. LREC},
  year         = {2020}
}

@misc{emilia,
  title        = {Emilia: An Extensive, Multilingual, and Diverse Speech Dataset for Large-Scale Speech Generation},
  author       = {He, Haorui and Shang, Zengqiang and Wang, Chaoren and Li, Xuan and Gu, Yicheng and Hua, Peiyu and Liu, Liwei and Yang, Chen and Li, Jiaqi and Shi, Peiyang and Wang, Yuancheng and Chen, Kai and Wu, Zhizheng},
  year         = {2024},
  eprint       = {2407.05361},
  archiveprefix = {arXiv},
  primaryclass = {cs.CL}
}

@inproceedings{gigaspeech,
  title        = {GigaSpeech: An Evolving, Multi-Domain ASR Corpus with 10,000 Hours of Transcribed Audio},
  author       = {Chen, Guoguo and Chai, Shuzhou and Wang, Guanbo and Du, Jiayu and Zhang, Wei-Qiang and Weng, Chao and Lu, Dan and Povey, Daniel and Trmal, Jan and Zhang, Junbo and Jin, Mingjie and Khudanpur, Sanjeev and Wang, Shinji and Wu, Shuai and Yang, Yong and Wang, Yaqing and Yu, Zhuo and Wang, Zeyu},
  booktitle    = {Proc. Interspeech},
  year         = {2021}
}

@inproceedings{kespeech,
  title        = {KeSpeech: An Open Source Speech Dataset of Mandarin and Its Eight Subdialects},
  author       = {Tang, Zhiyuan and Wang, Dong and Xu, Xiaoxue and Zheng, Hongbin and Lei, Yexin and Li, Jiawen and Zhang, Shuai and Zhu, Yifan and Meng, Jingjing and Li, Haoran and Xu, Xuyang and Zheng, Yuhang and Li, Shihao},
  booktitle    = {Proc. IEEE ASRU},
  year         = {2021}
}

@inproceedings{librispeech,
  title        = {LibriSpeech: An ASR Corpus Based on Public Domain Audio Books},
  author       = {Panayotov, Vassil and Chen, Guoguo and Povey, Daniel and Khudanpur, Sanjeev},
  booktitle    = {Proc. IEEE ICASSP},
  year         = {2015}
}

@misc{thchs30,
  title        = {THCHS-30: A Free Chinese Speech Corpus},
  author       = {Wang, Dong and Zhang, Xuewei},
  year         = {2015},
  eprint       = {1512.01882},
  archiveprefix = {arXiv},
  primaryclass = {cs.CL}
}

@inproceedings{wenetspeech,
  title        = {WenetSpeech: A 10,000+ Hours Multi-Domain Mandarin Corpus for ASR},
  author       = {Zhang, Bin and Lv, Hui and Guo, Pengcheng and Shao, Qijie and Yang, Chao and Xie, Lei and Xu, Xin and Bu, Hui and Chen, Xie and Zeng, Chuang and Wu, Di and Peng, Zhendong},
  booktitle    = {Proc. IEEE ICASSP},
  year         = {2022}
}

@inproceedings{tedlium,
  title        = {TED-LIUM: An Automatic Speech Recognition Dedicated Corpus},
  author       = {Rousseau, Anthony and Del{\'e}glise, Paul and Est{\`e}ve, Yannick},
  booktitle    = {Proc. LREC},
  year         = {2012}
}

@inproceedings{fleurs,
  title        = {FLEURS: Few-Shot Learning Evaluation of Universal Representations of Speech},
  author       = {Conneau, Alexis and Bapna, Ankur and Zhang, Yu and Ma, Min and von Platen, Patrick and Lozhkov, Anton and Cherry, Colin and Jia, Ye and Rivera, Clara and Kale, Mihir and Remez, Noam and Glavcev, Viktor and Gopala, Sanjay and Ni, Jian and Wu, Yu-Hsiang and Hsu, Po-Ning and Liu, Jason and Sahebi, Amir and Duquenne, Paul-Ambroise and Chen, Mia and Chau, Vinh and others},
  booktitle    = {Proc. IEEE SLT},
  year         = {2022}
}

\clearpage
\appendix
\section*{Appendix}
\phantomsection
\addcontentsline{toc}{section}{Appendix}
\setcounter{section}{1}
\setcounter{subsection}{0}

\subsection{Full Hyperparameter Configuration}
\label{sec:hyperparameters}
\begin{table}[!htb]
  \caption{Main 16-layer hyperparameters. The 14-layer run uses the same recipe and initializes from the recovered 16-layer checkpoint.}
  \label{tab:hparams}
  \centering
  \small
  \begin{tabular}{@{}lccc@{}}
    \toprule
    Parameter & Stage~0 & Stage~1 & Stage~2 \\
    \midrule
    Fraction / epochs & 0.05 epoch & 0.95 epoch & 1 epoch \\
    LR (audio tower) & $2\times10^{-5}$ & $2\times10^{-5}$ & $5\times10^{-6}$ \\
    LR (LoRA / tied head) & $10^{-4}/10^{-4}$ & $10^{-4}/10^{-4}$ & $5\times10^{-6}$/frozen \\
    Optimizer / weight decay & AdamW / 0.01 & AdamW / 0.01 & AdamW / 0.01 \\
    Per-device batch / accumulation & 8 / 2 & 8 / 2 & 8 / 2 \\
    Global batch size & 512 & 512 & 512 \\
    LoRA $r/\alpha$/dropout & 32 / 64 / 0.05 & 32 / 64 / 0.05 & 32 / 64 / 0.05 \\
    Temperature & 1.5 & 1.0 & -- \\
    $\lambda_{\rm layer}$ & 1.0 & 0.0 & -- \\
    $\lambda_{\rm bridge}$ & 1.0 & 0.5 & -- \\
    $\lambda_{\rm logit}$ & 0.2 & 0.1 & -- \\
    $\lambda_{\rm ce}$ & 0.5 & 1.0 & 1.0 \\
    On-policy start / fraction & -- & 0.2 / 0.2 & -- \\
    Union top-$k$ / weight mode & -- & 512 / none & -- \\
    Minimum / maximum new tokens & -- & 3 / 256 & -- \\
    Reject-batch threshold & -- & 0.5 & -- \\
    \bottomrule
  \end{tabular}
\end{table}

\subsection{Premature-EOS Safeguards}
\label{sec:eos}

Immediately after audio-encoder pruning, the model can emit EOS after only a few tokens, producing large deletion errors. Qwen3-ASR does not use a separate decoder cross-attention module: bridge outputs replace audio-placeholder embeddings in the causal decoder input. We hypothesize that pruning shifts these conditioning embeddings, weakening acoustic evidence for the pretrained decoder. We do not directly measure embedding mean/variance or EOS causality, so this account motivates the safeguards rather than constituting a mechanistic proof.

The implementation uses three safeguards. First, during main Stage~0/1 distillation, the tied lm\_head/token embedding is trainable at the decoder-side learning rate ($10^{-4}$; $5\times10^{-5}$ in the EOS ablation). Decoder Transformer weights remain frozen apart from LoRA. Second, on-policy generation enforces a fixed \texttt{min\_new\_tokens}$=3$; its maximum is duration-aware, $\min(256,\max(128,\lceil12d+32\rceil))$, where $d$ is the longest audio duration in the batch. Third, rollout filters reject budget-exhausted, implausibly long, and repetitive sequences. If the rejected fraction reaches 0.5, the scheduled on-policy microbatch falls back to teacher-forced distillation.

\begin{figure}[!htb]
    \centering
    \includegraphics[width=\textwidth]{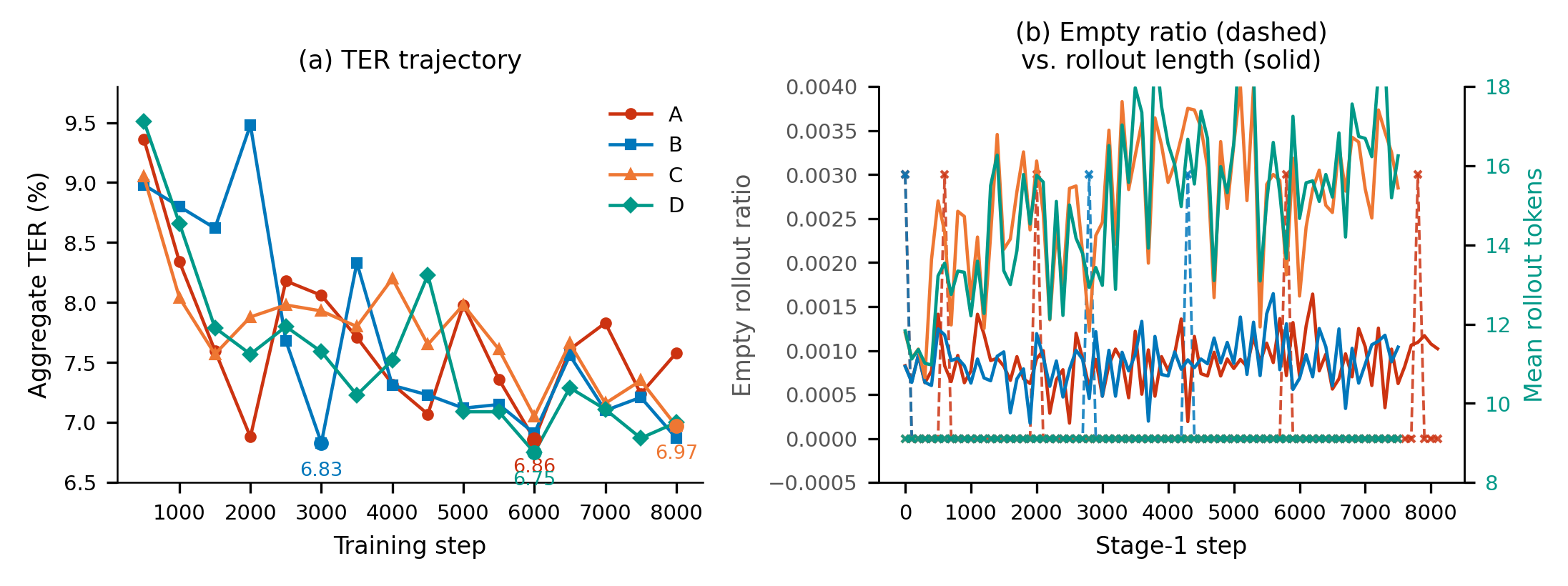}
    \caption{Premature-EOS ablation on prune-16. A disables both mitigations, B trains the tied head, C applies minimum-token gating, and D combines them. (a) Development TER. (b) Empty-rollout ratio (dashed) and mean rollout length (solid). All runs start from the same Stage~0 checkpoint.}
    \label{fig:eos}
\end{figure}

\begin{table}[!htb]
  \caption{Premature-EOS safeguards on prune-16 in a 2$\times$2 ablation. All configurations share the same Stage~0 checkpoint and Stage~1 recipe. Empty events count 100-step windows with at least one empty rollout; rejections sum budget, length, and repetition filters. TER is the best development-suite observation.}
  \label{tab:ablation_eos}
  \centering
  \small
  \begin{tabular}{@{}lccccc@{}}
    \toprule
    Configuration & lm\_head & min\_new & Empty events & TER & Rejections \\
    \midrule
    A: neither & frozen & 0 & 5 & 6.86 & 20 \\
    B: lm\_head & trainable & 0 & 3 & 6.83 & 11 \\
    C: gating & frozen & 3 & \textbf{0} & 6.97 & 2113 \\
    D: both & trainable & 3 & \textbf{0} & \textbf{6.75} & 2162 \\
    \bottomrule
  \end{tabular}
\end{table}

The no-mitigation condition has five windows with a nonzero empty ratio; the maximum is 0.3\% and the mean is 0.018\%. Training the tied output embedding reduces this to three events, while \texttt{min\_new\_tokens}$=3$ removes empty events in both gating conditions. Gating also converts some premature terminations into longer or degenerate rollouts: configurations C and D trigger 2113 and 2162 filter rejections, compared with 20 and 11 for A and B.

The best observed TER is 6.75\% for the combined configuration D. The tied-head-only configuration B reaches 6.83\% and shows only 0.04-pp best-to-last drift; A drifts from 6.86\% to 7.58\%. Mean rollout length is 15.06 tokens for D and 11.05 for A. These single runs support using the tied head and gating together in the main recipe, but the 0.08-pp B--D difference is too small to interpret without uncertainty estimates.

\subsection{Teacher-Scale Comparison}
\label{sec:teacher_ablation}

The self-teacher control uses the unpruned 18-layer Qwen3-ASR-0.6B model. The cross-scale condition uses Qwen3-ASR-1.7B and the required 2048$\rightarrow$1024 bottleneck projections. Other Stage~0/1 settings are matched, and both models are evaluated at their best development checkpoint before Stage~2.

\begin{table}[!htb]
  \caption{Teacher-scale comparison for the 16-layer Stage~1 student. Schedules, data, and optimization are matched; cross-scale training additionally uses the required hidden-width projections. Values are single-run best-checkpoint results.}
  \label{tab:teacher_scale}
  \centering
  \small
  \renewcommand{\arraystretch}{1.12}
  \begin{tabular}{@{}lccc@{}}
    \toprule
    Benchmark & Base (18L) & Self-teacher & Cross-scale \\
    \midrule
    AISHELL-1 (CER) & 3.33\% & 4.29\% & \textbf{3.30\%} \\
    Fleurs-zh (CER) & \textbf{2.80\%} & 4.17\% & 3.35\% \\
    Fleurs-en (WER) & \textbf{4.17\%} & 6.15\% & 4.28\% \\
    LibriSpeech test-clean (WER) & \textbf{2.48\%} & 6.00\% & 2.73\% \\
    THCHS-30 (CER) & \textbf{3.87\%} & 5.23\% & 4.10\% \\
    Tedlium (WER) & \textbf{3.35\%} & 11.28\% & 3.92\% \\
    LibriSpeech test-other (WER) & \textbf{5.39\%} & 9.89\% & 5.90\% \\
    CommonVoice v15 zh (CER) & 9.95\% & 11.71\% & \textbf{8.56\%} \\
    CommonVoice v15 en (WER) & 12.35\% & 14.89\% & \textbf{10.74\%} \\
    WenetSpeech-meeting (CER) & \textbf{8.36\%} & 10.85\% & 8.62\% \\
    \midrule
    \textit{Macro mean (\%)} & 5.61 & 8.45 & \textbf{5.55} \\
    \textit{Relative mean-error change (\%)} & -- & $+50.6$ & \textbf{$-1.1$} \\
    \bottomrule
  \end{tabular}
\end{table}

The cross-scale checkpoint is better on every public benchmark and lowers mean error from 8.45\% to 5.55\%. The largest gaps occur on Tedlium (11.28 vs.\ 3.92), LibriSpeech test-clean (6.00 vs.\ 2.73), and LibriSpeech test-other (9.89 vs.\ 5.90). Relative to the original baseline, the cross-scale model also improves CommonVoice zh/en and AISHELL-1, whereas the self-teacher is worse on all ten benchmarks. This pattern is consistent with useful cross-scale transfer, especially on English and higher-error conditions. Because the comparison has one seed and includes projection modules only when dimensions differ, we do not treat it as proof that the teacher creates capabilities absent from every unpruned student.

\subsection{Layer-Interaction Details}
\label{sec:interaction_detail}

\begin{table}[!htb]
  \caption{Pair probes for 16$\rightarrow$14 pruning after the same 0.3-epoch warm-up. TER is the macro mean over five development subsets, including proprietary SC. $\Delta$ is relative to $\{5,6\}$. Bold marks the best observed candidate.}
  \label{tab:ablation_selection}
  \centering
  \scriptsize
  \setlength{\tabcolsep}{4pt}
  \begin{tabular}{@{}llcccccc@{}}
    \toprule
    Candidate & Type & TER & $\Delta$ & AISHELL & CV-en & Fleurs-en & Wenet / SC \\
    \midrule
    $\{5,6\}$ & Adj. & \textbf{6.93} & -- & 0.63 & 16.97 & 6.85 & 7.30 / 2.91 \\
    $\{6,7\}$ & Adj. & 7.37 & +0.44 & 0.63 & 16.06 & 6.48 & 8.47 / 5.23 \\
    $\{8,9\}$ & Adj. & 7.75 & +0.82 & 0.63 & 23.39 & 5.37 & 6.42 / 2.91 \\
    $\{6,8\}$ & Non-adj. & 7.78 & +0.85 & 0.63 & 20.64 & 6.11 & 8.03 / 3.49 \\
    $\{3,6\}$ & Non-adj. & 8.12 & +1.19 & 0.63 & 17.89 & 6.48 & 8.03 / 7.56 \\
    $\{14,15\}$ & Adj. & 9.93 & +3.00 & 0.95 & 23.39 & 11.48 & 8.61 / 5.23 \\
    \bottomrule
  \end{tabular}
\end{table}

The single-layer results used to form the interaction comparison are L8=5.97\%, L6=6.11\%, and L5=6.15\%. Thus, the non-adjacent pair $\{6,8\}$ contains the two best constituent removals (mean 6.04\%) but reaches 7.78\% as a pair. The selected adjacent pair $\{5,6\}$ has a slightly worse constituent mean (6.13\%) but reaches 6.93\%. The descriptive interaction penalties are therefore 1.74 and 0.80 pp, respectively.

All three adjacent candidates in Table~\ref{tab:ablation_selection} rank above the two non-adjacent candidates, while $\{14,15\}$ confirms that adjacency alone is not sufficient. One possible account is that removing a contiguous sub-block creates one residual-stream discontinuity whereas dispersed removal creates two. This explanation is untested; causal activation analysis and a larger factorial candidate set are needed.

\subsection{Progressive versus Direct Pruning}
\label{sec:direct_pruning}

We compare the reported progressive path with a direct 18$\rightarrow$14 run that drops the same original layers $\{1,18,5,6\}$ and uses the same nominal recovery and data budget.

\begin{table}[!htb]
  \caption{Direct 18$\rightarrow$14 pruning versus progressive 18$\rightarrow$16$\rightarrow$14 pruning. Both remove $\{1,18,5,6\}$ and use the same recovery/data budget. Single-run best-checkpoint results; bold marks the best observed value.}
  \label{tab:ablation_direct}
  \centering
  \small
  \renewcommand{\arraystretch}{1.12}
  \begin{tabular}{@{}lccc@{}}
    \toprule
    Benchmark & Base & Direct 18$\rightarrow$14 & Progressive \\
    \midrule
    AISHELL-1 (CER) & \textbf{3.33\%} & 3.81\% & 3.39\% \\
    Fleurs-zh (CER) & \textbf{2.80\%} & 3.76\% & 3.32\% \\
    Fleurs-en (WER) & \textbf{4.17\%} & 5.94\% & 5.10\% \\
    LibriSpeech test-clean (WER) & 2.48\% & 3.42\% & \textbf{2.45\%} \\
    THCHS-30 (CER) & \textbf{3.87\%} & 4.62\% & 4.17\% \\
    Tedlium (WER) & \textbf{3.35\%} & 5.10\% & 3.95\% \\
    LibriSpeech test-other (WER) & \textbf{5.39\%} & 6.31\% & 5.52\% \\
    CommonVoice v15 zh (CER) & 9.95\% & 9.79\% & \textbf{8.36\%} \\
    CommonVoice v15 en (WER) & \textbf{12.35\%} & 15.22\% & 12.49\% \\
    WenetSpeech-meeting (CER) & \textbf{8.36\%} & 9.29\% & 8.78\% \\
    \midrule
    \textit{Macro mean (\%)} & \textbf{5.61} & 6.73 & 5.75 \\
    \bottomrule
  \end{tabular}
\end{table}

Progressive pruning reaches 5.75\% mean error, compared with 6.73\% for direct pruning, and is better on all ten benchmarks. The largest direct-minus-progressive gaps are CommonVoice en (+2.73 pp), CommonVoice zh (+1.43 pp), and Tedlium (+1.15 pp). This shows that the progressive path is preferable under the tested budget. It does not establish that no alternative schedule or larger budget could improve direct pruning.

\subsection{On-Policy Strategy Comparison}
\label{sec:opd_ablation}

\begin{figure}[!htb]
    \centering
    \includegraphics[width=0.94\textwidth]{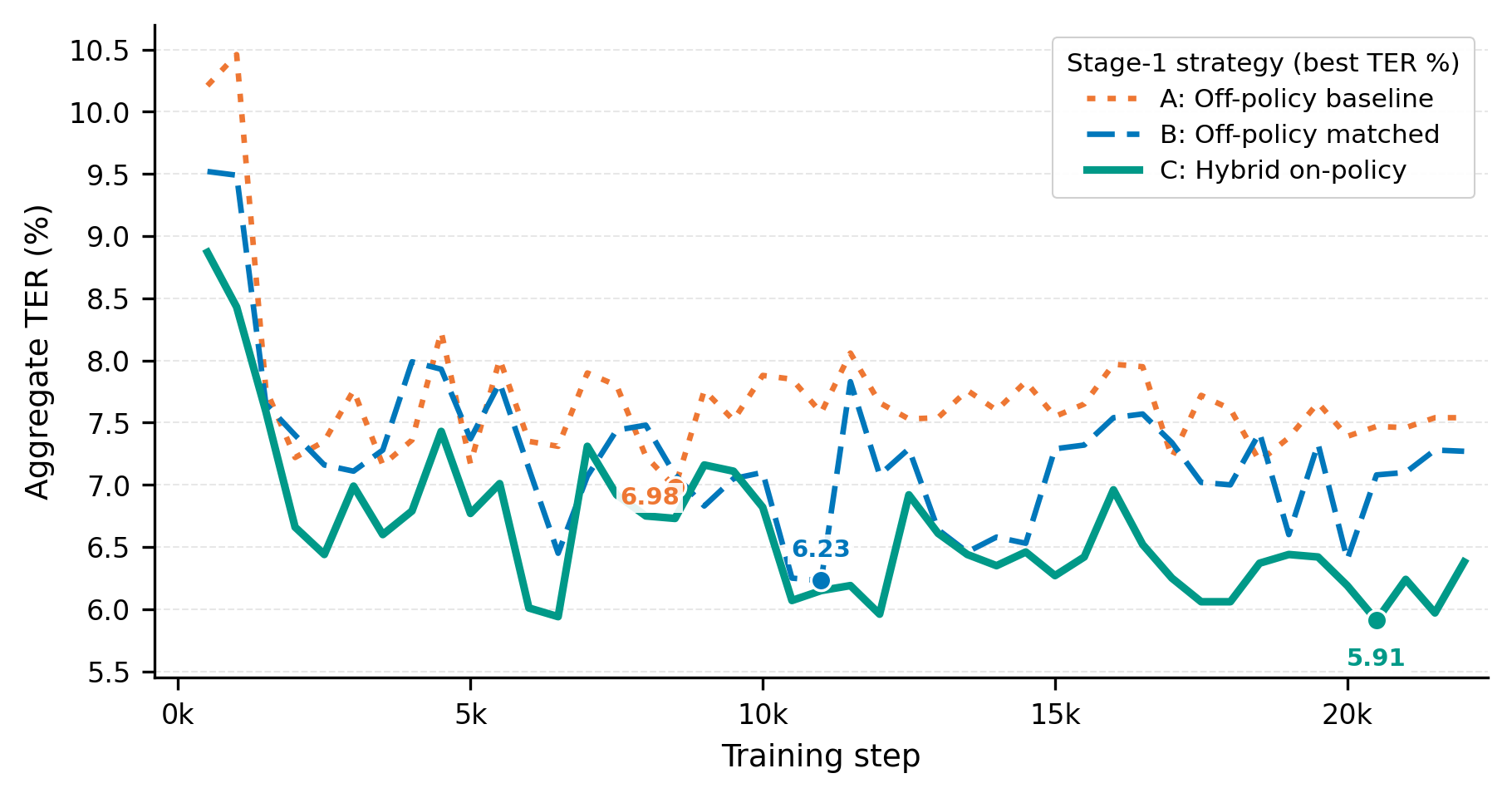}
    \caption{Stage~1 strategy dashboard for plain off-policy, matched off-policy, and hybrid on-policy training; lower TER is better. Exact best and final values in the accompanying text are taken from the archived checkpoint logs.}
    \label{fig:opd_strategy}
\end{figure}

Figure~\ref{fig:opd_strategy} compares three Stage~1 runs on prune-16: a plain off-policy baseline, an off-policy control matched to the extra student forward and loss settings used on scheduled steps, and hybrid on-policy training. The dashboard provides the visual trajectory; numerical comparisons use the archived checkpoint logs.

The best observed TER values are 6.98\% for plain off-policy, 6.23\% for matched off-policy, and 6.41\% for hybrid on-policy. Final observations are 7.43\%, 6.51\%, and 6.99\%, respectively. Hybrid training therefore improves over the plain baseline but does not outperform the matched off-policy control in this experiment. We retain the hybrid recipe in the main run because it directly supervises student-generated contexts and works well in the complete pipeline, while recognizing that this ablation does not establish accuracy superiority. More seeds and a sweep over rollout fraction are needed.

\subsection{Inference Efficiency}
\label{sec:efficiency}

\begin{table}[!htb]
  \caption{Descriptive inference measurements averaged over more than 50 utterances of varying lengths. No run-to-run variance was retained. Peak device memory includes the full model, so encoder-only pruning changes it modestly.}
  \label{tab:efficiency}
  \centering
  \small
  \begin{tabular}{@{}lrrrr@{}}
    \toprule
    Model & Encoder (ms) & End-to-end (ms) & RTF & Peak memory (MB) \\
    \midrule
    \multicolumn{5}{c}{In-vehicle PPU} \\
    Baseline 18L & 14 & 486 & 0.0786 & 1500.3 \\
    Prune-14 & 11 & 463 & 0.0748 & 1434.4 \\
    Relative change & $-21.4\%$ & $-4.7\%$ & $-4.8\%$ & $-4.4\%$ \\
    \addlinespace
    \multicolumn{5}{c}{NVIDIA H800} \\
    Baseline 18L & 88 & 1672 & 0.256 & 1500.3 \\
    Prune-14 & 78 & 1629 & 0.245 & 1458.2 \\
    Relative change & $-11.4\%$ & $-2.6\%$ & $-4.3\%$ & $-2.8\%$ \\
    \bottomrule
  \end{tabular}
\end{table}

The 14-layer model reduces encoder time by 21.4\% on the in-vehicle PPU and 11.4\% on H800. End-to-end reductions are 4.7\% and 2.6\%, because autoregressive decoding dominates total time. These measurements support encoder pruning as a localized efficiency improvement, especially when encoder and decoder are pipelined, but not as a large end-to-end speedup by itself.

\end{document}